\documentclass{aa}

\usepackage[final]{epsfig} 
\usepackage[final]{graphics}

\usepackage[american,german,english]{babel}

\usepackage{natbib}
\bibpunct{(}{)}{;}{}{}{,}

\usepackage{latexsym}
\usepackage{amssymb}

\begin{document}

\def\nuInu{$\nu$I$_{\nu}$}
\def\dnuInu{$\delta$($\nu$I$_\nu$)}
\def\nWmsr{nWm$^{-2}$sr$^{-1}$}

\title{Sky Confusion Noise in the Far-Infrared: Cirrus, Galaxies and
       the Cosmic Far--Infrared Background%
  \thanks{Based on observations with ISO, an ESA project with instruments
    funded by ESA Member States (especially the PI countries: France,
    Germany, the Netherlands and the United Kingdom) and
    with the participation of ISAS and NASA.}
  }
                   
\author{ Cs.~Kiss\inst{1,2}
  \and  P.~\'Abrah\'am\inst{1,2} 
  \and  U.~Klaas\inst{1} 
  \and  M.~Juvela\inst{3}
  \and  D.~Lemke\inst{1}}
\offprints{Cs.~Kiss, pkisscs{@}konkoly.hu}
\institute{
  Max-Planck-Institut f\"ur Astronomie, K\"onigstuhl~17,
     D-69117~Heidelberg, Germany
  \and  Konkoly Observatory of the Hungarian Academy of Sciences, 
    P.O. Box 67, H-1525~Budapest, Hungary 
  \and  Helsinki University Observatory, T\"ahtitorninm\"aki,
        P.O. Box 14, SF--00014 University of Helsinki, Helsinki  
   }

\date{Compiled: \today ---- Received / Accepted ...}

\titlerunning{Sky Confusion Noise in the Far-Infrared}
\abstract{
We examined the sky confusion noise in 40 sky regions by analysing 
175 far-infrared (90--200\,$\mu$m) maps obtained with ISOPHOT, 
the photometer on-board the Infrared Space Observatory. 
For cirrus fields with
$\rm \langle \rm B \rangle > 5$\,MJysr$^{-1}$ the 
formula based on IRAS data (Helou~\&~Beichman, \cite{Helou+Beichman_90}) 
predicts confusion noise values within a factor of 2
to our measurements.
The dependence of the sky confusion noise on the 
surface brightness was determined for the wavelength range 
90\,$\le \lambda \le$\,200\,$\mu$m.
We verified that the confusion noise 
scales as N\,$\propto \langle \rm B \rangle ^{1.5}$, independent
of the wavelength and confirmed 
N\,$\propto \lambda^{2.5}$ for $\rm \lambda \ge 100$\,$\mu$m.
The scaling of the noise value at different separations 
between target and reference positions was investigated 
for the first time, providing a practical formula. 
Since our results confirm the applicability of the 
{Helou~\&~Beichman}~\cite{Helou+Beichman_90} formula,
the cirrus confusion noise predictions made for future space missions 
with telescopes of a similar size can be trusted. 
{ At 90 and 170\,$\mu$m a noise term with a Poissonian spatial 
distribution was detected in the faintest fields 
($\rm \langle B \rangle \le 3-5\,MJysr^{-1}$), which we interpret
as fluctuations in the Cosmic Far-Infrared Background (CFIRB).
Applying ratios of the fluctuation amplitude to the absolute level 
of 10\% and 7\% at 90 and 170\,$\mu$m, respectively, as supported by 
model calculations, 
we achieved a new simultaneous determination of the fluctuation
amplitudes and the surface brightness of the CFIRB.
The fluctuation amplitudes are 7$\pm$2\,mJy and
15$\pm$4\,mJy at 90 and 170\,$\mu$m, respectively.
We obtained a CFIRB surface brightness of B$_0$\,=\,0.8$\pm$0.2\,MJysr$^{-1}$
(\nuInu\,=\,14$\pm$3\,\nWmsr) 
at 170\,$\mu$m and an upper limit of 1.1\,MJysr$^{-1}$ 
(\nuInu\,=\,37\,\nWmsr) at 90$\mu$m.}
}
\maketitle

\section{Introduction \label{introduction}}

Sky confusion noise causes an uncertainty in the determination
of the source flux, due to the 
variation of the sky brightness between the 
target (on-source) and reference (off-source) 
positions. This noise cannot be overcome by longer integration times,
thus it constitutes a basic limitation for the detection of
very faint point sources. 

At far-infrared wavelengths the two major components 
responsible for the sky confusion noise are dust emission
from irregularly shaped interstellar clouds, the ``galactic cirrus''
\citep[Low~et~al.,][]{Low_84}, and a component of the 
Extragalactic Background built up from the accumulated light of 
faint unresolved galaxies along the line of sight 
\citep[Guiderdoni~et~al.,][]{Guiderdoni_97}. 
     
The confusion noise due to cirrus can be characterized by a 
formalism applied first by 
{Gautier~et~al.}~\cite{Gautier_92} for the
IRAS 100\,$\mu$m scans. They computed the second order structure function
for a far-infrared map : 
\begin{equation} 
\rm 
S({\theta}) = \Big\langle \Big| B(x) - {{B( x-{\theta} ) + B( x+{\theta} )}\over{2}}
\Big| ^2 \Big\rangle _x
\label{strucfunc}
\end{equation}
where $\rm B$ is the sky brightness, $\rm x$ is the location of the
target, $\rm \theta$ is the separation between the target and reference 
sky positions and the average is taken in spatial coordinates over the 
whole map. 
The noise due to sky brightness fluctuations, N, is defined as:
\begin{equation} 
\rm 
N(\theta) = \sqrt{ S({\theta}) } \times {\Omega}
\label{noisedef}
\end{equation}    
where $\rm {\Omega}$ is the solid angle of 
the measuring aperture.

\begin{figure}[h!]
\begin{minipage}{85mm}
\centerline{\epsfig{file=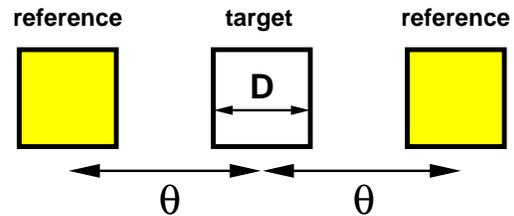, width=6.7cm}}
\end{minipage}
\caption{Measurement configuration to compute the structure noise.
{Gautier~et~al.}~\cite{Gautier_92} and 
{Helou~\&~Beichman}~\cite{Helou+Beichman_90} derived their 
results for the resolution limit $\rm \theta_{min} = 2\,D$.
In Sect.~\ref{knoise} we also consider the cases of 
$\rm \theta = k{\cdot}D, k=2...7$ }
\label{fig:aperture}
\end{figure}

{ The measured confusion noise depends both on the size of the 
measuring aperture D and on the angular separation  $\theta$ 
between the on-source and off-source positions. However, if the sky
brightness distribution does not show any characteristic scale length 
(e.g. random fluctuations or fractal structure), the only important 
factor is the $\theta$/D ratio.}
{ For their study of the IRAS 100\,$\mu$m maps 
{Gautier~et~al.}~\cite{Gautier_92} 
defined a standard measurement configuration}
where the on-source position is bracketed by two off-source 
positions separated by $\pm\,\theta$ (Fig.~\ref{fig:aperture}) 
{ and the ratio is fixed to the resolution (Nyquist) limit
$\theta$\,=\,$\rm\theta_{min}$\,=\,2D.}

{ {{Gautier~et~al.}~\cite{Gautier_92}} pointed out that the 
Fourier power spectra of the IRAS 100$\mu$m scans were generally well 
represented by a power law, and that the  
structure noise can be linked to the parameters of the  
power spectra by the following relation: 
\begin{equation} 
\rm 
N~{\propto}~ \Big( {d\over{d_0}} \Big) ^{1-{{\alpha}\over{2}}} \times P_0^{1\over{2}} 
\label{noise+fourier}
\end{equation}      
{ where $\rm d^{-1}$ is the spatial frequency, $\rm d_0^{-1}$ is a 
reference spatial frequency, P$_0$ is the Fourier 
power at d$^{-1}_0$, and $\alpha$ is the spectral index of the 
Fourier power spectrum. It was also found that most scans can be
described by the empirical relationships $\alpha$\,$\approx$\,3 and
$\rm P_0 \propto {\langle}B{\rangle}^3$. 

Based on the IRAS results 
{Helou~\&~Beichman}~\cite{Helou+Beichman_90} (H\&B)
proposed a practical formula to predict the cirrus confusion noise,
assuming that the empirical relationships of 
{Gautier~et~al.}~\cite{Gautier_92}
are valid at other wavelengths, too.
They also took into account
that at the resolution limit of the telescope
$\rm d/d_0$ can be expressed by the resolution parameter 
of Fraunhofer diffraction, $\rm\lambda/D_t$, where $\lambda$ is the 
wavelength of the measurement and $\rm D_t$ is the diameter
of the telescope primary mirror. Under these terms they expressed
the cirrus confusion noise as}}:   
\begin{equation} 
\rm 
{{N_{H{\&}B}}\over{1\,mJy}} = 0.3 {\times} \Big( {{\lambda}\over{100\,{\mu}m}}
    \Big) ^{2.5}
   \Big( {{D_t}\over{1\,m}} 
   \Big) ^{-2.5}  \Big( {{\langle B_{\lambda} \rangle}\over{1\,MJysr^{-1}}}
   \Big) ^{1.5}
\label{NHB}
\end{equation}     
{ Here $\rm B_{\lambda}$ is the surface brightness at the wavelength 
of the observation. 
The formula shows that the confusion noise depends on the wavelength in
two separate ways: (1) via the variation of the surface brightness 
(spectral energy distribution of the emitting medium) and (2) via 
the dependence of the resolution parameter on $\lambda$.} 
This relation { which is valid for the standard measurement 
configuration as defined in Fig.~\ref{fig:aperture}} has been widely used 
for preparing far-infrared space telescope observations, and applied
even for other configurations. 
The first noise analysis of four ISOPHOT maps was performed
by {Herbstmeier~et~al.}~\cite{Herbstmeier}, 
{ carrying out the first high spatial resolution study of the}
galactic cirrus at $\lambda{>}$100$\mu$m. 
They proved that the ISOPHOT sensitivity at 
far-infrared wavelengths is limited by sky confusion noise
rather than by instrumental uncertainties.     

The confusion noise caused by the CFIRB 
has different properties from the cirrus confusion noise.
It can be represented by a Poissonian spatial distribution, therefore
the confusion noise value is { independent} of the 
separation between the target and the reference positions. 
{Ackermann~et~al.}~\cite{Ackermann92} investigated the sensitivity limits,
including the effect of the CFIRB, for 
ISOPHOT and predicted that the long wavelength 
observations would be limited by galaxy confusion in the 
regions of faintest cirrus.   
Recently {Lagache~\&~Puget}~\cite{Lagache+Puget_2000},
{Matsuhara~et~al.}~\cite{Matsuhara} and Juvela et al. 
(for an overview see {Lemke~et~al.}~\cite{Lemke2000})
analysed ISOPHOT maps and separated the cirrus
and extragalactic components. They announced the 
detection of the CFIRB with a 
fluctuation power in the range of $\rm 5-12{\times}10^3\,Jy^2sr^{-1}$
at 170 and 180\,$\mu$m, close to the predictions of 
{Guiderdoni~et~al.}~\cite{Guiderdoni_97}.  
This leads to a galaxy confusion noise limit of 13--22\,mJy 
for ISOPHOT at this wavelength.

For this paper we analysed 175
suitable far-infrared maps from the ISO Archive. 
Our goals were (1) to test the applicability of
the {Helou~\&~Beichman}~\cite{Helou+Beichman_90} formula, especially at
$\lambda\,>$\,100\,$\mu$m; (2) to determine the relative 
contributions of the galactic cirrus and the extragalactic 
components in the faintest regions of the sky;    
(3) to derive an easy-to-use 
formalism for predicting the total sky confusion noise 
for instruments on ISO, SIRTF and HERSCHEL and for various 
measurement configurations, and { (4) to determine 
the Cosmic Far-Infared Background Radiation via its fluctuations.}
        
\section{Observations and data analysis\label{observations}}

\subsection{Selection of ISOPHOT maps\label{isophot_maps}}

We selected 175 maps from the ISO Archive 
\citep[Kessler~et~al.,][]{Kessler2000},
observed in the 90 to 200\,${\mu}$m wavelength range,
covering 40 different sky regions.
All maps were obtained with the 
ISOPHOT instrument \citep[Lemke~et~al.,][]{Lemke_96} 
on-board the ISO satellite \citep[Kessler~et~al.,][]{Kessler_96}, 
in the PHT22 raster observing mode
\citep[Laureijs~et~al.,][]{Laureijs2000}. 
Our selection criteria were that the maps should be  
larger than 5$\times$5 raster points, corresponding to
$\sim$\,8$'$\,$\times$\,8$'$ for the C100 
(3$\times$3 pixel array, 46\arcsec ~pixel size) and C200 
(2$\times$2 pixel array, 92\arcsec ~pixel size) detectors.
Some maps were observed with full oversampling, 
i.e. the field was redundantly covered without gaps between
detector pixels.  
The final map images are a composite of the individual array images
on the different raster positions. The size of the image pixels 
is always a detector pixel.   

We excluded maps with obvious individual structures 
(resolved stars, galaxies, planetary nebulae, etc.). 
The number of selected maps is
94, 4, 65 and 12 using the C1\_90, C1\_100, C2\_170 and C2\_200 filters,
respectively, and they cover a wide range of surface brightnesses 
($\sim$\,1--100\,MJysr$^{-1}$).
A more detailed description of the 
maps will be presented in a forthcoming paper 
(Kiss et al. 2001, in prep.) dealing with the spatial structure 
of the extended emission. 

\subsection{Data reduction\label{data_reduction}}
Basic data reduction was performed using the ISOPHOT Interactive
Analysis software, PIA~V8.2 
\citep[Gabriel~et~al.,][]{Gabriel97}
 with the standard batch mode set-up.
Flat-fielding had to be applied, which corrects for the 
residual responsivity differences 
of the individual detector pixels. { We chose the 
First Quartile Normalization method, which uses the first 
quartiles of the maps' brightness distribution, computed for 
each detector pixel, for normalization.
The reliability of this statistical flat-field method was tested 
in the case of oversampled maps, were redundant observations of the
same sky position by each detector pixel are available. These 
tests confirmed that the First Quartile Normalization gives 
accurate flat-field values, and 
was therefore adopted for this analysis.}

{ The maps still contain the zodiacal light (ZL) emission, 
and a contribution from the CFRIB. The zodiacal emission
does not contribute to the sky confusion noise, because
its distribution is smooth on the scale 
of our maps \citep[\'Abrah\'am~et~al.,][]{Abraham97}.
Therefore, we can eliminate it from the total brightness. 
First, we determined the non color corrected ratio of the ZL--to--total 
emission at the DIRBE wavelengths, using the 
COBE/DIRBE weekly maps and the COBE/DIRBE sky and zodi atlas 
({Hauser~et~al.}\cite{Hauser98}, {Kelsall~et~al.}~\cite{Kelsall})}.
{ We colour corrected the ZL component adopting a spectral energy 
distribution (SED) shape of a black body with a temperature of 
270\,K (Leinert et al. \cite{Leinert} 
find 255\,K\,$\le$\,T$\rm _{ZL}$\,$\le$\,$297\,K$).
The difference between the uncorrected total and ZL emission
was color corrected by assigning a cirrus SED to it, 
assuming a 20\,K modified black body with $\rm\nu^2$ emissivity law.
Then interpolating the corrected ratios in between the COBE/DIRBE 
wavelengths, we estimated the ratios of the ZL-to-total emission for 
the ISOPHOT filter central wavelengths. The absolute ZL contribution in the 
ISOPHOT maps was computed by multiplying this ratio with the average total 
brightness measured by ISOPHOT 
\citep[see also H\'eraudeau~et~al.,][]{Heraudeau}.} 

{ The remaining surface brightnesses include the CFIRB and the 
cirrus emission. In Sect.~\ref{results} we 
calculate the sky confusion noise and 
in Sect.~\ref{EGB} we separate the cirrus and Extragalactic Background
fluctuations, and derive the value of the CFIRB.}

\subsection{Instrument noise\label{instrumental_noise}}

The determination of the instrument noise is crucial in order to get the
real value of the sky confusion noise. 
In addition to the basic noise components
(read-out noise, dark current variations, etc.), we also include
in the instrument noise  
uncertainties related to the generation of the final maps (flat-field).  
 
There are several ways to determine the instrument noise. 
The fundamental difference between them 
is the time scale over which the noise estimation was taken.
We examined the following four methods:
\begin{itemize}
 \item \underline{Ramp-noise:}
       Each measurement is composed of many individual integrations 
       on each raster position. A set of non-destructive read-outs
       builds up an integration ramp whose slope provides the 
       signal of this integration 
       \citep[Lemke~et~al.,][]{Lemke_96}.   
       From each measurement we created two maps using the even and 
       odd integration ramps separately.  
       The instrument noise was calculated from the difference 
       between the 'even' and 'odd' maps.
       This noise estimate provides information on the detector
       stability on a time scale of seconds, the typical difference 
       between adjacent ramps.  
  
\item \underline{PIA-noise:}
      PIA provides a { determination of the signal uncertainty 
      of the individual raster positions, which can also be 
      adopted as an estimate of N$_{\rm inst}$.
      It reflects the error propagation 
      of random error components in
       (1) the linear fitting of the integration ramps,
      and (2) the averaging of all signals per raster point}
      \footnote{ The on-line PIA-manual is available at the 
      URL: ``http://www.iso.vilspa.esa.es/manuals/PHT/pia/pia.html''}.
      The time scale here is the time spent on an individual 
      raster position, typically in the order of a minute. 
\item \underline{Flat-field-noise:}
       This noise value was calculated from 
       { the variation of the brightness at the same sky position 
       in maps created from individual detector pixels
       (i.e. in case of pixel redundancy), after correction 
       by the appropriate flat-fielding factors. }
       { Since these factors are kept constant for the whole 
       map, although the detector pixels show different time dependent
       transient curves, this instrument 
       noise estimate includes the effects of the long term 
       changes in the detector behaviour.}        
       Therefore the time scale is the observational time of the 
       whole map, typically in the order of an hour.       
       { Since the transient behaviour depends on the illumination
       level, the flat-field-noise may also depend on the brightness 
       of the map.}       
\item \underline{Repetition-noise:}
       If a sky region has been observed several times, then it 
       is possible to derive a noise value from the comparison 
       of the individual maps. The typical time scale here is 
       the difference in the observation dates.
       { It provides a measure of the measurement reproducibility.}
\end{itemize}

The ramp- and PIA-noise could be calculated for each map, 
but the flat-field-noise could be evaluated only in maps with 
full oversampling (see Sect.~\ref{isophot_maps}) and the
repetition-noise was only computed for one particular sky region
{(Marano\,1 field)} consisting of several submaps,
each observed four times.
A comparison of the different noise estimates showed that the
ramp-noise provided the lowest N$\rm{_{inst}}$ values, typically 
a factor of 2 lower than the PIA-noise. 
The flat-field noise was higher than the PIA one, and the repetition 
noise was found to be nearly identical to the flat-field-noise.
Since in the case of the structure noise calculation 
{(Eq.~\ref{strucfunc})} the target and reference 
positions were observed at times typically separated by several minutes,
the correct instrument noise should lie between the PIA- and the
flat-field-noise values. To be on the safe side, we chose the value of the 
flat-field-noise, although it may represent a somewhat 
conservative estimate, which might cause a slight underestimation 
in the final confusion noise values. Since this noise value is not available 
for each map, we { took} the uncertainties provided by PIA 
and scaled them according to the { mean} ratio of 1.35 and 
1.65 { of the flat-field- and PIA-noise values}
for the C100 and C200 detectors, respectively.

\begin{figure}[h!]
\epsfig{file=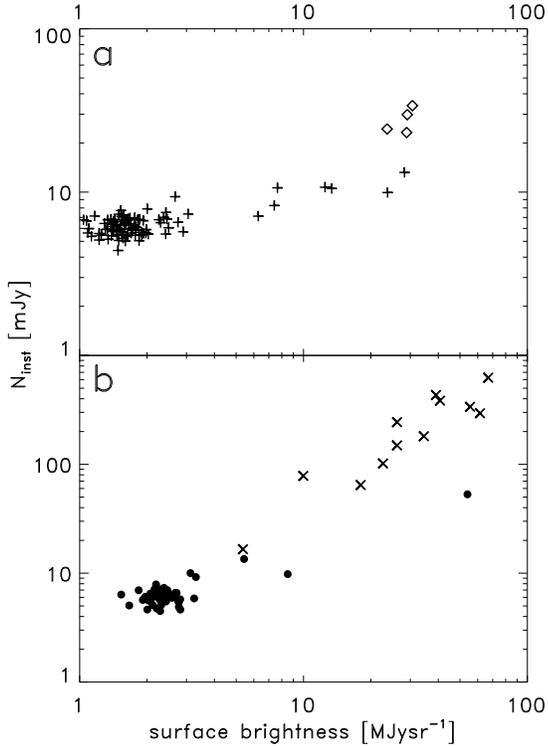, width=8.5cm}
\caption{
Instrument noise values measured on the ISOPHOT maps.
{ In the case of oversampled maps the flat-field noise, 
otherwise the scaled up PIA-noise was adopted. The mean scaling
factor between the PIA- and the flat-field noise values are 1.35 and
1.65 for the C\_100 and C\_200 detectors, respectively.}
({\bf a}) C100 detector: C1\_90 and C1\_100 filters (plus signs and 
diamonds, respectively),
({\bf b}) C200 detector: C2\_170 and C2\_200 filters 
(black dots and crosses, respectively)}
\label{fig:n_instr}
\end{figure}

The typical instrument noise values we found (see Fig.~\ref{fig:n_instr})
are somewhat higher than that of other authors, 
e.g. a factor of 2 compared to {Lagache~et~al.}~\cite{Lagache+Puget_2000}. 
This discrepancy can be traced back to the co-addition of four 
independent maps in their case, reducing the final noise values by  
this factor. 
  
{ Following 
{Herbstmeier~et~al.}~\cite{Herbstmeier} we assumed that} the sky brightness 
fluctuations and instrument noise contributions are statistically 
independent, therefore the measured structure noise can be expressed as:
\begin{equation}
\rm 
N_{meas}^2 \le N^2 + 2 \cdot N_{inst}^2
\label{eq:nmeas}
\end{equation}
where $\rm N_{meas}$ is the measured structure noise, 
$\rm N$ is the real structure noise and $\rm N_{inst}$ is the
instrument noise.  
{ We used this formula for the subtraction of the instrument 
noise from the measured noise, by assuming equality between the two 
sides of the expression. In order to test this assumption 
we compared   
the confusion noise values of four C2\_170 maps, where
the same sky region was mapped repeatedly.
The confusion noise must be the same and
only the instrument noise values can differ in repeated observations.
Assuming equality 
in Eq.~\ref{eq:nmeas} we obtained identical confusion noise
values for all four measurements and for any separations.}


\section{Results \label{results}}

\subsection{Confusion noise at the resolution limit \label{noiselimit}}
\begin{figure}
\resizebox{8.5cm}{!}{\includegraphics{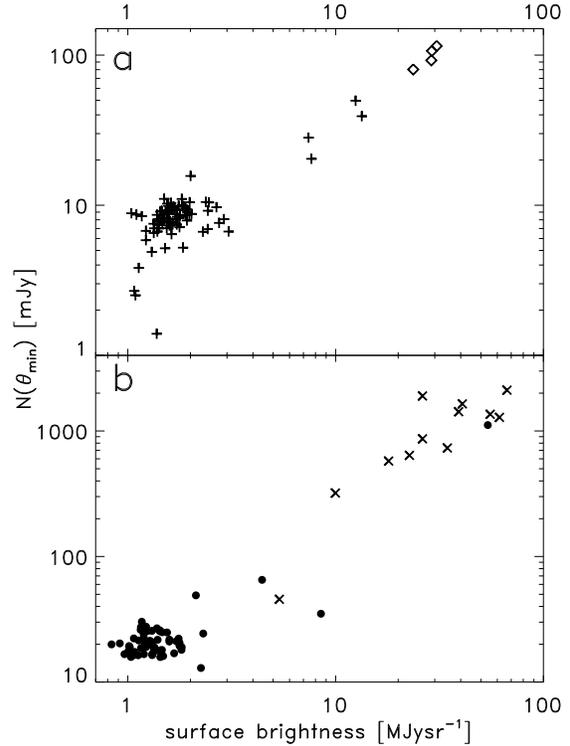}}
\caption{Confusion noise at the resolution limit
$\rm \theta_{min}$ vs. surface brightness of the fields. 
({\bf a}) ISOPHOT C100 filters,  C1\_90 and C1\_100. 
({\bf b}) ISOPHOT C200 filters,  C2\_170 and C2\_200.
The symbols are the same as in Fig.~\ref{fig:n_instr}}
\label{fig:bigfig}
\vspace{0.3cm}
\end{figure}

Fig.~\ref{fig:bigfig} presents our results on the sky confusion noise
(after removing the instrument noise using Eq.~\ref{eq:nmeas}) 
measured with the four ISOPHOT filters 
C1\_90, C1\_100, C2\_170 and C2\_200. 
{ We used the convention $\theta$\,=\,2D, defined by 
{Gautier~et~al.}~\cite{Gautier_92} (see Sect.~\ref{introduction})}. 
Following the H\&B formula (Eq.~\ref{NHB}) we
fitted the data points for each filter separately, 
assuming a power-law relationship
between the confusion noise and the average brightness of the 
field, but also allowing a constant factor: 
\begin{equation}
\rm {{N(\theta_{min})}\over{1\,mJy}} = C_0 + {C_1}{\times}
\Big( {{\langle \rm B \rangle}\over{1\,MJysr^{-1}}} \Big) ^{\eta}
\label{eq:fit}
\end{equation}    
The coefficients C$_0$, C$_1$ and $\eta$ 
are listed in Table~\ref{table:fit}. 
In the case of the C1\_100 and C2\_200 filters
no appropriate determination of the C$_0$ offset was possible 
due to the lack of faint regions.
Eq.~\ref{eq:fit} can be used to predict the confusion noise 
for the ISOPHOT filters studied here.  
Our confusion noise values measured at 170\,$\mu$m are in
agreement with a value of 45\,mJy obtained by 
{Dole~et~al.}~\cite{Dole2001} in the 
FIRBACK regions, taking
into account that their value is a source flux confusion limit, 
which differs from ours by the footprint-fraction and is therefore 
a factor of 2 higher.       
\begin{table}
\small
\begin{tabular}{ccccc}
\hline
filter      & C1\_90      & C1\_100     & C2\_170     & C2\_200     \\ \hline
$\rm C_0$   & 6.6$\pm$1.9   & --          & 10.2$\pm$2.5  & --          \\ 
$\rm C_1$   & 0.7$\pm$0.3 & 0.6$\pm$0.1 & 3.0$\pm$0.3   & 4.4$\pm$0.2 \\ 
$\rm \eta$  & 1.55$\pm$0.12 & 1.53$\pm$0.28 & 1.47$\pm$0.11 & 1.56$\pm$0.19 \\ 
$\rm \theta_{min}$ & 92\arcsec & 92\arcsec & 184\arcsec & 184\arcsec \\ 
$\Omega$ ($\rm 10^{-7}$sr)  & 0.6469 & 0.7030 & 2.6438 & 2.8120 \\
\hline 
\end{tabular}
\normalsize
\caption{Results of the fits using the measured values 
of the sky confusion noise at the resolution limit
 $\rm N(\theta_{min})$ and 
the average surface brightness of the field $\langle B \rangle$
(see also Eq.~\ref{eq:fit} and the text for details).
Separation limit and effective solid angle values are also presented.
In the case of the C1\_100 and C2\_200 filters
no appropriate determination of the C$_0$ offset was possible 
due to the lack of faint regions.}
\label{table:fit}
\end{table}

\subsection{Comparison with the H\&B formula \label{comp-hb}}

\begin{figure}[!]
\resizebox{8.5cm}{!}{\includegraphics{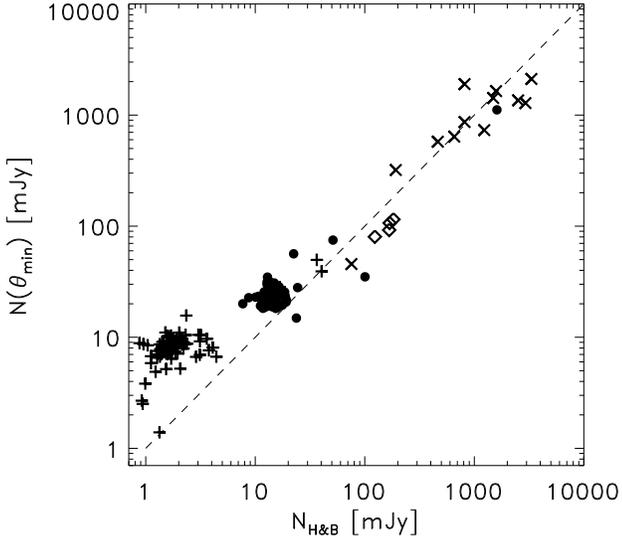}}
\caption[]{Confusion noise at the resolution limit 
determined by this study as compared to the prediction of
{Helou~\&~Beichman}~\cite{Helou+Beichman_90}. 
The symbols are the same as in Fig~\ref{fig:n_instr}.}
\label{fig:hbcomp}
\end{figure}
  
In Fig.~\ref{fig:hbcomp} we compare our results with the 
predictions of the H\&B formula (Eq.~\ref{NHB}). 
In the surface brightness range where the cirrus emission
dominates (5--30\,MJysr$^{-1}$) the H\&B formula predicts confusion noise 
values within a factor of 2 of our results. 
The scatter cannot be attributed to measurement uncertainties, and 
probably it reflects the physical 
differences among the fields.   
For the brightest fields the H\&B formula seems to 
systematically overestimate the measured confusion noise values.
Such regions, however, contain molecular clouds, therefore they may have
different spatial structure from a cirrus region and should be
described by a different power law.      
The figure also shows a large 
discrepancy at low surface brightness which could well be 
treated by the introduction of an offset (see above).
A possible explanation of the deviations in the faint range
is discussed in Sect.~\ref{EGB}.  

\begin{figure}
\resizebox{8.5cm}{!}{\includegraphics{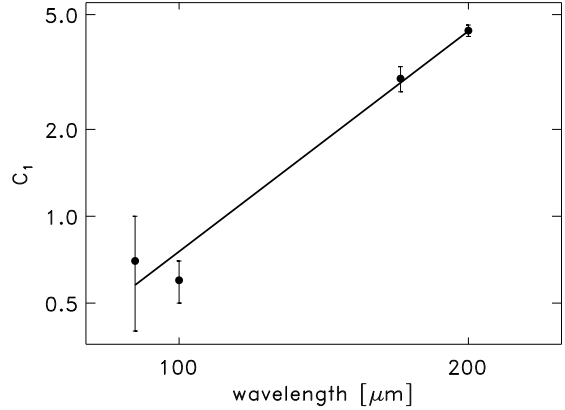}}
\caption{C$_1$ coefficients of the fits in Table~\ref{table:fit}
vs. the appropriate wavelength (logarithmic-logarithmic plot). }
\label{fig:lambdaexp}
\end{figure}

Although the H\&B formula gives rather good predictions 
for the surface brightness range it was developed for
(5--30\,MJysr$^{-1}$), its coefficients, especially the
three exponents, rely partly on the analysis of IRAS 100$\mu$m
data, partly on theoretical considerations. 
Our data set offers a unique chance to test 
some of these exponents for the first time. 
The dependence of the confusion noise on the surface brightness
was described by H\&B using a power law with an exponent of 1.5
(Eq.~\ref{NHB}), independently of the wavelength.
This exponent $\eta$ was determined by our fits for four 
wavelengths in Sect.~\ref{noiselimit}. The measured values in
Table~\ref{table:fit} are very close to the value of 1.5, 
proving the wavelength independence of this exponent.

{ In the next step we tested the dependence of the confusion noise 
on the resolution parameter $\rm \lambda/D_t$}, also
assumed to be in the form of a power law with an exponent of 2.5
by H\&B (Eq.~\ref{NHB}).
{ Since the diameter of the telescope mirror $\rm D_t$ is the same 
regardless of the wavelength, in Fig.~\ref{fig:lambdaexp} we plotted 
the wavelength dependence of the C$_1$ coefficient from 
Table~\ref{table:fit}}. 
The figure confirms the assumption of a 
power law, and gives an exponent of 2.53$\pm$0.31, in  
agreement with the value in Eq.~\ref{NHB}. 
The dependence of the confusion noise on the diameter of the 
telescope (Eq.~\ref{NHB}) was not possible to check, since 
the telescope mirror of ISO had the same size as that of IRAS.
Our results verify exponents of the H\&B formula 
for the first time by observations over a wide range.   

\subsection{Confusion noise for larger separations \label{knoise}}

The H\&B formula provides information only on the measurement 
configuration presented in Fig.~\ref{fig:aperture}, at the
resolution limit $\rm \theta_{min}$. However, 
in many applications the beam separation may be different.
In most ISOPHOT measurements the observer was allowed 
to specify any separation up to 3\arcmin. 
Staring on-off measurements and sparse maps allowed even
larger separations \citep[Laureijs~et~al.,][]{Laureijs2000}. 
Properties of the sky confusion noise for these 
cases were not investigated so far. 

\begin{figure}[!]
\resizebox{8.5cm}{!}{\includegraphics{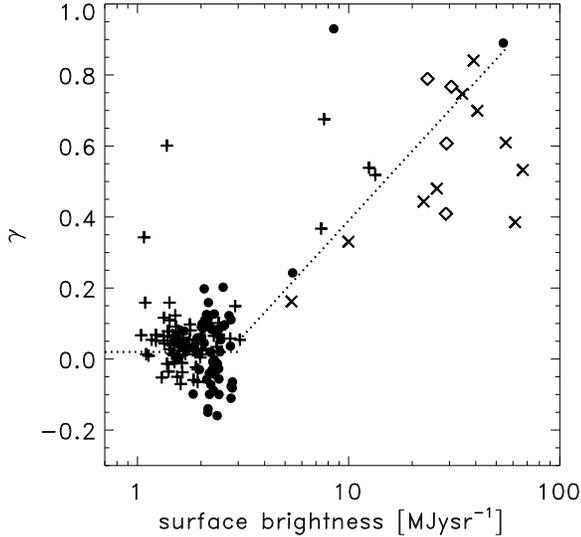}}
\caption{$\gamma$ values of the 
$\rm N(q\cdot\theta_{min}) = N(\theta_{min}){\cdot}q^{\gamma}$
fits for four ISOPHOT filters versus the surface brightness 
of the fields (see the text for details). The symbols are
the same as in Fig.~\ref{fig:n_instr}} 
\label{fig:knoise}
\end{figure}

As indicated in Sect.~\ref{introduction}, the ratio 
of the confusion noise values measured at different 
separations is sensitive to the spatial distribution of the 
dominant component of the emission (galactic cirrus or 
distant galaxies). 
We found that the dependence of the sky confusion
noise on the separation between target and reference 
sky positions can be described by a simple power law:
\begin{equation}
\rm N(q\cdot\theta_{min}) = 
  N(\theta_{min}) \times q^{\gamma}
\label{eq:knoise}
\end{equation}
where q=1,1$1\over 2$...3$1\over 2$ and
$\rm \gamma$ is constant for a specific map.
We fitted $\rm \gamma$ for all maps. 
The results are presented in Fig.~\ref{fig:knoise}, 
which suggests that the $\gamma$ value is a 
filter-independent parameter of the field, depending only 
on the surface brightness. 
For faint fields $\gamma$ is 
close to zero, while it increases strongly with increasing 
surface brightness. We found that the behaviour 
can be approximated by the following relation (dotted curve
in Fig.~\ref{fig:knoise}):  
\begin{equation}
\rm \gamma =  \left \{ \begin{array}{lr}
                        0.02  & 
	         \rm \,\,\,\,\, if \,\,\,{\langle \rm B \rangle} \le \rm 3\,MJysr^{-1}\\
  0.65{\times}{\rm log_{\rm 10}} {\langle \rm B \rangle} - 0.26  
                          & \rm \,\,\,\,\,if \,\,\,{\langle \rm B \rangle} > 3\,MJysr^{-1}  
			\end{array}
			 \right.
\label{eq:gamma}
\end{equation}
Eq.~\ref{eq:knoise} and \ref{eq:gamma} together with 
Eq.~\ref{eq:fit} and the coefficients in Table~\ref{table:fit}
provide a practical tool for the estimation of the confusion noise
for a large variety of measurements.    
 

\section{Discussion \label{discussion}}


\subsection{Variations of $\huge \gamma$ \label{gamma}}

As shown in Sect.~\ref{knoise}, for low surface brightness fields 
the $\gamma$ values are close to zero. 
In those regions the noise distribution differs from that 
expected for the galactic cirrus 
\citep[see e.g. Ackermann~et~al.,][]{Ackermann92}), 
originating from its multifractal 
structure. $\gamma\approx$\,0 is typical for a Poissonian
distribution.
 
Higher $\gamma$ values in brighter regions may be caused by the
cirrus structure. 
Despite this general trend the scatter of $\gamma$ 
at higher surface brightness is relatively large and 
there are a few regions measured by the C1\_90 filter with 
high $\gamma$ values at low brightness. 
These discrepancies are probably caused by  
real physical differences between the fields.
The $\gamma$ parameter depends on the spectral index 
of the Fourier power spectrum of the field 
(cf. Eq.~\ref{noise+fourier}).                
This relationship will be analysed in detail based on experimental data
in a forthcoming paper (Kiss et al. 2001, in prep.). 
Differences in physical properties may originate in 
different chemical/dust composition, spatial structure, 
gas-to-dust or molecular-to-neutral gas ratio.


\subsection{Fluctuations due to the Extragalactic Background \label{EGB}}
\subsubsection{Findings from this work}
When fitting the sky confusion noise as a function of the surface brightness 
in Sect.~\ref{noiselimit} a constant term C$_0$ was allowed for.
At both 90 and 170\,$\mu$m definite positive values were obtained at the
3--4$\sigma$ significance level. 
Assuming that the cirrus confusion noise follows the power law
behaviour also at very low surface brightness, this offset 
cannot be attributed to cirrus.
The spatial distribution of the confusion noise below 
$\sim$3\,MJysr$^{-1}$ has different characteristics than in brighter
cirrus fields (see Sect.~\ref{knoise} and Fig.~\ref{fig:knoise}).
The likely origin of this noise component, as suggested by its
invariability with the surface brightness and its Poissonian
spatial distribution, would be the fluctuation due to the 
Cosmic Far-Infrared Background (CFIRB). 

If this interpretation is correct, than Eq.~\ref{eq:fit} 
provides a new method to determine the value of the CFIRB fluctuation.
The features of this method are (1) the subtraction 
of the cirrus component via its dependence on the surface 
brightness and (2) that this dependence is calibrated simultaneously 
on the brighter cirrus fields of the same database.  
An advantage of this method is the utilization of the 
largest ISOPHOT database for the determination of the CFIRB fluctuations. 
The results are not sensitive to characteristics in individual fields. 
On the other hand, any uncertainty in the
surface brightness calibration of the cirrus would introduce an uncertainty in
the fluctuation value, too. 

The values given by the fitting procedure in Sect.~\ref{noiselimit}
for the C$\rm_0$ coefficients are 6.6$\pm$1.9 and
10.2$\pm$2.5\,mJy at 90 and 170\,$\mu$m, respectively.
{ In this fitting the surface brightness values still 
included a contribution from the CFIRB itself. Subtraction of the CFIRB 
increases the fluctuation amplitudes. 
Varying the CFIRB surface 
brightness B$_0$ within the assumed range of 0.1 to 1.5\,MJysr$^{-1}$
and subtracting it from the total surface brightness   
we used Eq.~\ref{eq:fit} to derive the corresponding C$_0$ coefficients 
which were then transferred to surface brightness fluctuations 
$\delta$B$_0$ as $\delta$B$_0$\,=\,C$\rm _0$/$\Omega_{\lambda}$, 
where $\Omega_{\lambda}$ is the effective solid angle of a detector pixel
(see Table~\ref{table:fit}).
 
The results are shown in Fig.~\ref{fig:CFIRB}.
As it can be seen in this figure, the dependence of $\delta$B$_0$
on the assumed value of B$_0$ is not very strong, 
therefore, regardless of the real value of the CFIRB, a well confined 
value of $\delta$B$_0$ can be derived. 
We obtained C$_0$\,=\,7$\pm$2\,mJy 
($\delta$B$_0$\,=\,0.11$\pm$0.03\,MJysr$^{-1}$) at 90$\mu$m and
C$_0$\,=\,15$\pm$4\,mJy 
($\delta$B$_0$\,=0.06\,$\pm$0.02\,MJysr$^{-1}$) at 
170$\mu$m.

In most models CFIRB fluctuations are caused by galaxy clustering, 
and, due to the flat power spectrum, the $\delta$B$_0$/B$_0$ 
value is constant in the 1\degr\, to 5\arcmin\, resolution range
({Kashlinsky et al.}~\cite{Kashlinsky}, {Haiman \& Knox}~\cite{Haiman}).
At finer resolution (higher spatial frequencies) effects by individual
galaxies have to be considered. An estimate of  
$\delta$B$_0$/B$_0$ (or \dnuInu/(\nuInu)) can be performed
following {Bond et al.}~\cite{Bond}, assuming biased galaxy clustering.
Using eq.~6.35 from {Bond et al.}~\cite{Bond}, the fluctuation ratio is : 
$\rm \delta(\nu I_\nu)/(\nu I_\nu) = 0.05 \times 
( 1\arcmin / \sigma )^{0.4}$
where $\sigma$ is the smoothing angle, 
and the beam profile is approximated by a Gaussian.
Equations 6.24 and 7.2 in Bond et al.~\cite{Bond} give 
$\sigma$\,=\,12\arcsec\, and 23\arcsec\, for 90 and 170\,$\mu$m,
respectively, for ISOPHOT. From this results 
\dnuInu/(\nuInu)\,=\,10\% at 90\,$\mu$m and 
\dnuInu/(\nuInu)\,=\,7\% at 170\,$\mu$m. 
\begin{figure}
\epsfig{file=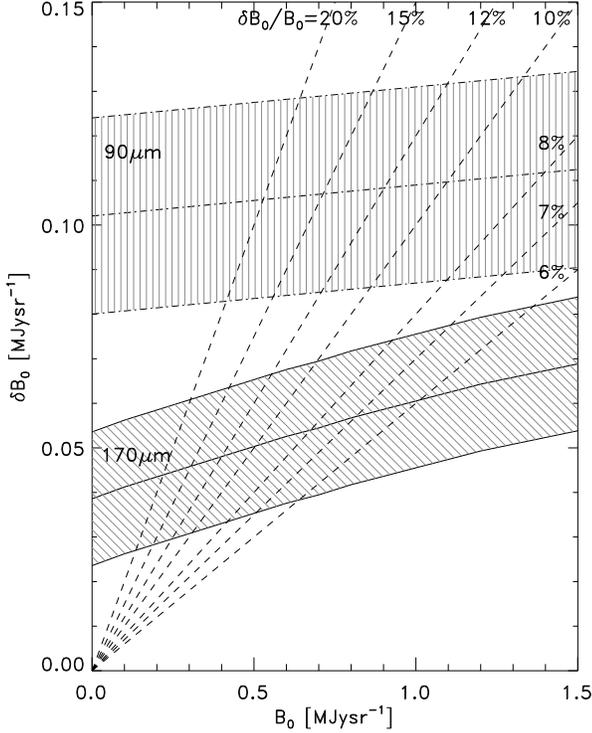, width=8.5cm}
\caption[]{ CFIRB fluctuation amplitudes versus the
CFIRB absolute level assumed.  
The hatched regions represent the formal errors of the fits in 
Eq.~\ref{eq:fit} for 90 and 170\,$\mu$m. A set of dashed lines marks  
$\delta$B$_0$/B$_0$ = 6, 7, 8, 10, 12, 15 and 20\%.}
\label{fig:CFIRB}
\end{figure}  

{ At 170\,$\mu$m $\delta$B$_0$/B$_0$\,=\,7\% provides 
B$_0$\,=\,0.8$\pm$0.2\,MJysr$^{-1}$ equivalent to
\nuInu\,=\,14$\pm$3\,\nWmsr.  

At 90\,$\mu$m the dependence of $\delta$B$_0$ on B$_0$ is even 
weaker than at 170\,$\mu$m, therefore a well defined value of 
0.11$\pm$0.03\,MJysr$^{-1}$ can be determined. 
The B$_0$ value of 1.1$\pm$0.3\,MJysr$^{-1}$,
equivalent to \nuInu\,=\,37$\pm$10\,\nWmsr), 
is the same as the average surface brightness of the faintest 90\,$\mu$m 
maps in our data set. The expected cirrus contribution in these fields 
-- predicted from the $\sim$0.8\,MJysr$^{-1}$ value at 170\,$\mu$m by using 
a typical cirrus spectrum 
(modified black-body SED, $\rm\nu^2$ emissivity law, 20\,K temperature)
-- is about 0.3\,MJysr$^{-1}$.  
This result suggests that the 1.1\,MJysr$^{-1}$ is an upper limit
and the most likely value of the CFIRB lies at the lower end of 
the uncertainty range, i.e. $\sim$0.8\,MJysr$^{-1}$ 
equivalent to \nuInu\,=\,30\,\nWmsr.

Our results satisfy the three main criteria of a CFIRB detection
as proposed by {Hauser~\&~Dwek}~\cite{Hauser+Dwek}:
(1) We achieved a detection of a positive value of the
fluctuation amplitudes (and therefore the absolute value as well) 
at 90 and 170\,$\mu$m at the 4$\sigma$ level.
(2) We removed the contribution by all known foreground 
(noise) emitters (zodiacal light, instrument noise, galactic cirrus). 
The properties of the spatial distribution related to the 
remaining noise term agree with that expected for the CFIRB fluctuations. 
(3) Due to the statistical nature of our method, the detection of the 
same well defined positive constant term in different areas of the 
far-infrared sky indicates the isotropy of this component. 
}}
{
\subsubsection{Comparison with other results}
The 170\,$\mu$m confusion noise C$_0$\,=\,15$\pm$4\,mJy
is in good agreement with the 13\,mJy obtained by Juvela et al. 
(see the overview of {Lemke~et~al.}~\cite{Lemke2000}), and somewhat 
lower than the values of 18\,mJy determined by 
{Lagache~\&~Puget}~\cite{Lagache+Puget_2000} 
from the analysis of the Marano\,1 field and the 22\,mJy derived
by {Matsuhara~et~al.}~\cite{Matsuhara} from the 
investigation of the Lockman Hole region.  
Our derived CFIRB value of 14$\pm$3\,\nWmsr~ is the same as 
predicted from models by {Pei et al.}~\cite{Pei}. 
Also the COBE measurements calibrated on the FIRAS photometric
scale 
\citep[Hauser et al.][~for an overview see Hauser \& Dwek 2001]{Hauser98}
give quite similar levels (15$\pm$6\,\nWmsr~ at 140$\mu$m and 
13$\pm$2\,\nWmsr~ at 240$\mu$m) despite the big difference in 
spatial resolution. 

The 90\,$\mu$m confusion noise C$_0$\,=\,7$\pm$2\,mJy
is close (although somewhat lower) to the 11\,mJy 
found by {Matsuhara~et~al.}~\cite{Matsuhara}. 
With regard to the 90\,$\mu$m absolute value
{Schlegel~et~al.}~\cite{Schlegel98} provided the same 1.1\,MJysr$^{-1}$
value as an upper limit for the CFIRB at 100\,$\mu$m, 
and also {Lagache et al.}~\cite{Lagache2000} and 
{Finkbeiner et al.}~\cite{Finkbeiner} reported a detection of a
100\,$\mu$m CFIRB level of 23$\pm$6\,\nWmsr~
and 25$\pm$8\,\nWmsr~, respectively, which are close to our
upper limit. 
}
\subsection{Confusion limits for far-infrared space
                       telescopes \label{predictions}}

After the comparison of instrument and confusion noise
values (see Fig.~\ref{fig:n_instr} and \ref{fig:bigfig}) 
it is obvious that even 
in the faintest sky regions the instrument noise is below the 
sky confusion noise by a factor of 2--3 for the C2\_170 filter.
This confirms the results of {Herbstmeier~et~al.}~\cite{Herbstmeier}, 
showing that the ISOPHOT C200 detector was limited 
by sky brightness fluctuations rather than by instrumental effects. 
In the case of the C1\_90 filter the instrumental and confusion noise 
are similar in strength at the resolution limit.
Since our results confirm the applicability of the H\&B formula,
the predictions made for other space missions
with telescopes of a similar size (SIRTF, Astro-F) 
on the basis of this formula can be trusted.
Predictions made for the HERSCHEL satellite (3.6m mirror), however,
still have to partially rely on assumptions, since we were not able to test 
the dependence of the confusion noise on the diameter of the telescope 
primary mirror. 
It remains to be demonstrated that the spatial structure of cirrus 
does not change below the resolution limit of ISO.     

\section{Summary \label{summary}}

Using measurements obtained with ISOPHOT  
we investigated the properties of the sky 
confusion noise on a large sample of maps in the 
90\,$\le$\,$\lambda$\,$\le$\,200\,$\mu$m wavelength range.
We described the dependence of the sky confusion noise on the 
surface brightness for four selected ISOPHOT filters.
We verified that the confusion noise 
scales as N\,$\propto \langle \rm B \rangle ^{1.5}$
for the resolution limit, independent
of the wavelength. We also confirmed that, due to the
dependence on the resolution parameter,
N\,$\propto \lambda^{2.5}$ at $\rm \lambda \ge 100$\,$\mu$m.
According to our results for cirrus fields with
$\rm \langle \rm B \rangle > 5$\,MJysr$^{-1}$ the 
{Helou~\&~Beichman}~\cite{Helou+Beichman_90}
formula predicts confusion noise values within a factor of 2.
The scaling of the noise value at different separations 
between target and reference positions was investigated 
for the first time, providing a useful formula to estimate 
the confusion for different separations, too. 
{ At 90 and 170\,$\mu$m a noise term with a Poissonian spatial 
distribution was detected in the faintest fields 
($\rm \langle B \rangle \le 3-5\,MJysr^{-1}$), which we interpret
as fluctuations in the Cosmic Far-Infared
Background (CFIRB). 
With a ratio of the fluctuation amplitude to the absolute level 
of 10\% and 7\% at 90 and 170\,$\mu$m, respecively, 
we determined the fluctuation
amplitudes and the surface brightness of the CFIRB.
The fluctuation amplitudes are 7$\pm$2\,mJy and
15$\pm$4\,mJy at 90 and 170\,$\mu$m, respectively.
We obtained a CFIRB surface brightness of B$_0$\,=\,0.8$\pm$0.2\,MJysr$^{-1}$
(\nuInu\,=\,14$\pm$3\,\nWmsr) 
at 170\,$\mu$m and an upper limit of 1.1\,MJysr$^{-1}$ 
(\nuInu\,=\,37\,\nWmsr) at 90$\mu$m.}

\begin{acknowledgements}
\sloppy
The development and operation of ISOPHOT were supported by MPIA and
funds from Deut\-sches Zentrum f\"ur Luft- und Raumfahrt 
(DLR, formerly DARA). The ISOPHOT Data Center at MPIA is supported
by Deut\-sches Zentrum f\"ur Luft- und Raumfahrt e.V. (DLR) with
funds of Bundesministerium f\"ur Bildung und Forschung, 
grant~no.~50~QI~9801~3. This research was partly supported by the
ESA PRODEX programme (No.~14594/00/NL/SFe) and
by the Hungarian Research Fund (OTKA F-022566).
M.J. acknowledges the support of the Academy of 
Finland Grant no.~1011055.
The authors are responsible for the contents of this publication.

\end{acknowledgements}



\end{document}